\documentclass[a4paper]{jpconf}
\usepackage{graphicx}
\usepackage{hyperref}
\usepackage{xcolor}
\usepackage{fancyvrb}
\usepackage{color}

\makeatletter
\def\PY@reset{\let\PY@it=\relax \let\PY@bf=\relax%
    \let\PY@ul=\relax \let\PY@tc=\relax%
    \let\PY@bc=\relax \let\PY@ff=\relax}
\def\PY@tok#1{\csname PY@tok@#1\endcsname}
\def\PY@toks#1+{\ifx\relax#1\empty\else%
    \PY@tok{#1}\expandafter\PY@toks\fi}
\def\PY@do#1{\PY@bc{\PY@tc{\PY@ul{%
    \PY@it{\PY@bf{\PY@ff{#1}}}}}}}
\def\PY#1#2{\PY@reset\PY@toks#1+\relax+\PY@do{#2}}

\expandafter\def\csname PY@tok@gd\endcsname{\def\PY@tc##1{\textcolor[rgb]{0.63,0.00,0.00}{##1}}}
\expandafter\def\csname PY@tok@gu\endcsname{\let\PY@bf=\textbf\def\PY@tc##1{\textcolor[rgb]{0.50,0.00,0.50}{##1}}}
\expandafter\def\csname PY@tok@gt\endcsname{\def\PY@tc##1{\textcolor[rgb]{0.00,0.27,0.87}{##1}}}
\expandafter\def\csname PY@tok@gs\endcsname{\let\PY@bf=\textbf}
\expandafter\def\csname PY@tok@gr\endcsname{\def\PY@tc##1{\textcolor[rgb]{1.00,0.00,0.00}{##1}}}
\expandafter\def\csname PY@tok@cm\endcsname{\let\PY@it=\textit\def\PY@tc##1{\textcolor[rgb]{0.25,0.50,0.50}{##1}}}
\expandafter\def\csname PY@tok@vg\endcsname{\def\PY@tc##1{\textcolor[rgb]{0.10,0.09,0.49}{##1}}}
\expandafter\def\csname PY@tok@vi\endcsname{\def\PY@tc##1{\textcolor[rgb]{0.10,0.09,0.49}{##1}}}
\expandafter\def\csname PY@tok@mh\endcsname{\def\PY@tc##1{\textcolor[rgb]{0.40,0.40,0.40}{##1}}}
\expandafter\def\csname PY@tok@cs\endcsname{\let\PY@it=\textit\def\PY@tc##1{\textcolor[rgb]{0.25,0.50,0.50}{##1}}}
\expandafter\def\csname PY@tok@ge\endcsname{\let\PY@it=\textit}
\expandafter\def\csname PY@tok@vc\endcsname{\def\PY@tc##1{\textcolor[rgb]{0.10,0.09,0.49}{##1}}}
\expandafter\def\csname PY@tok@il\endcsname{\def\PY@tc##1{\textcolor[rgb]{0.40,0.40,0.40}{##1}}}
\expandafter\def\csname PY@tok@go\endcsname{\def\PY@tc##1{\textcolor[rgb]{0.53,0.53,0.53}{##1}}}
\expandafter\def\csname PY@tok@cp\endcsname{\def\PY@tc##1{\textcolor[rgb]{0.74,0.48,0.00}{##1}}}
\expandafter\def\csname PY@tok@gi\endcsname{\def\PY@tc##1{\textcolor[rgb]{0.00,0.63,0.00}{##1}}}
\expandafter\def\csname PY@tok@gh\endcsname{\let\PY@bf=\textbf\def\PY@tc##1{\textcolor[rgb]{0.00,0.00,0.50}{##1}}}
\expandafter\def\csname PY@tok@ni\endcsname{\let\PY@bf=\textbf\def\PY@tc##1{\textcolor[rgb]{0.60,0.60,0.60}{##1}}}
\expandafter\def\csname PY@tok@nl\endcsname{\def\PY@tc##1{\textcolor[rgb]{0.63,0.63,0.00}{##1}}}
\expandafter\def\csname PY@tok@nn\endcsname{\let\PY@bf=\textbf\def\PY@tc##1{\textcolor[rgb]{0.00,0.00,1.00}{##1}}}
\expandafter\def\csname PY@tok@no\endcsname{\def\PY@tc##1{\textcolor[rgb]{0.53,0.00,0.00}{##1}}}
\expandafter\def\csname PY@tok@na\endcsname{\def\PY@tc##1{\textcolor[rgb]{0.49,0.56,0.16}{##1}}}
\expandafter\def\csname PY@tok@nb\endcsname{\def\PY@tc##1{\textcolor[rgb]{0.00,0.50,0.00}{##1}}}
\expandafter\def\csname PY@tok@nc\endcsname{\let\PY@bf=\textbf\def\PY@tc##1{\textcolor[rgb]{0.00,0.00,1.00}{##1}}}
\expandafter\def\csname PY@tok@nd\endcsname{\def\PY@tc##1{\textcolor[rgb]{0.67,0.13,1.00}{##1}}}
\expandafter\def\csname PY@tok@ne\endcsname{\let\PY@bf=\textbf\def\PY@tc##1{\textcolor[rgb]{0.82,0.25,0.23}{##1}}}
\expandafter\def\csname PY@tok@nf\endcsname{\def\PY@tc##1{\textcolor[rgb]{0.00,0.00,1.00}{##1}}}
\expandafter\def\csname PY@tok@si\endcsname{\let\PY@bf=\textbf\def\PY@tc##1{\textcolor[rgb]{0.73,0.40,0.53}{##1}}}
\expandafter\def\csname PY@tok@s2\endcsname{\def\PY@tc##1{\textcolor[rgb]{0.73,0.13,0.13}{##1}}}
\expandafter\def\csname PY@tok@nt\endcsname{\let\PY@bf=\textbf\def\PY@tc##1{\textcolor[rgb]{0.00,0.50,0.00}{##1}}}
\expandafter\def\csname PY@tok@nv\endcsname{\def\PY@tc##1{\textcolor[rgb]{0.10,0.09,0.49}{##1}}}
\expandafter\def\csname PY@tok@s1\endcsname{\def\PY@tc##1{\textcolor[rgb]{0.73,0.13,0.13}{##1}}}
\expandafter\def\csname PY@tok@ch\endcsname{\let\PY@it=\textit\def\PY@tc##1{\textcolor[rgb]{0.25,0.50,0.50}{##1}}}
\expandafter\def\csname PY@tok@m\endcsname{\def\PY@tc##1{\textcolor[rgb]{0.40,0.40,0.40}{##1}}}
\expandafter\def\csname PY@tok@gp\endcsname{\let\PY@bf=\textbf\def\PY@tc##1{\textcolor[rgb]{0.00,0.00,0.50}{##1}}}
\expandafter\def\csname PY@tok@sh\endcsname{\def\PY@tc##1{\textcolor[rgb]{0.73,0.13,0.13}{##1}}}
\expandafter\def\csname PY@tok@ow\endcsname{\let\PY@bf=\textbf\def\PY@tc##1{\textcolor[rgb]{0.67,0.13,1.00}{##1}}}
\expandafter\def\csname PY@tok@sx\endcsname{\def\PY@tc##1{\textcolor[rgb]{0.00,0.50,0.00}{##1}}}
\expandafter\def\csname PY@tok@bp\endcsname{\def\PY@tc##1{\textcolor[rgb]{0.00,0.50,0.00}{##1}}}
\expandafter\def\csname PY@tok@c1\endcsname{\let\PY@it=\textit\def\PY@tc##1{\textcolor[rgb]{0.25,0.50,0.50}{##1}}}
\expandafter\def\csname PY@tok@o\endcsname{\def\PY@tc##1{\textcolor[rgb]{0.40,0.40,0.40}{##1}}}
\expandafter\def\csname PY@tok@kc\endcsname{\let\PY@bf=\textbf\def\PY@tc##1{\textcolor[rgb]{0.00,0.50,0.00}{##1}}}
\expandafter\def\csname PY@tok@c\endcsname{\let\PY@it=\textit\def\PY@tc##1{\textcolor[rgb]{0.25,0.50,0.50}{##1}}}
\expandafter\def\csname PY@tok@mf\endcsname{\def\PY@tc##1{\textcolor[rgb]{0.40,0.40,0.40}{##1}}}
\expandafter\def\csname PY@tok@err\endcsname{\def\PY@bc##1{\setlength{\fboxsep}{0pt}\fcolorbox[rgb]{1.00,0.00,0.00}{1,1,1}{\strut ##1}}}
\expandafter\def\csname PY@tok@mb\endcsname{\def\PY@tc##1{\textcolor[rgb]{0.40,0.40,0.40}{##1}}}
\expandafter\def\csname PY@tok@ss\endcsname{\def\PY@tc##1{\textcolor[rgb]{0.10,0.09,0.49}{##1}}}
\expandafter\def\csname PY@tok@sr\endcsname{\def\PY@tc##1{\textcolor[rgb]{0.73,0.40,0.53}{##1}}}
\expandafter\def\csname PY@tok@mo\endcsname{\def\PY@tc##1{\textcolor[rgb]{0.40,0.40,0.40}{##1}}}
\expandafter\def\csname PY@tok@kd\endcsname{\let\PY@bf=\textbf\def\PY@tc##1{\textcolor[rgb]{0.00,0.50,0.00}{##1}}}
\expandafter\def\csname PY@tok@mi\endcsname{\def\PY@tc##1{\textcolor[rgb]{0.40,0.40,0.40}{##1}}}
\expandafter\def\csname PY@tok@kn\endcsname{\let\PY@bf=\textbf\def\PY@tc##1{\textcolor[rgb]{0.00,0.50,0.00}{##1}}}
\expandafter\def\csname PY@tok@cpf\endcsname{\let\PY@it=\textit\def\PY@tc##1{\textcolor[rgb]{0.25,0.50,0.50}{##1}}}
\expandafter\def\csname PY@tok@kr\endcsname{\let\PY@bf=\textbf\def\PY@tc##1{\textcolor[rgb]{0.00,0.50,0.00}{##1}}}
\expandafter\def\csname PY@tok@s\endcsname{\def\PY@tc##1{\textcolor[rgb]{0.73,0.13,0.13}{##1}}}
\expandafter\def\csname PY@tok@kp\endcsname{\def\PY@tc##1{\textcolor[rgb]{0.00,0.50,0.00}{##1}}}
\expandafter\def\csname PY@tok@w\endcsname{\def\PY@tc##1{\textcolor[rgb]{0.73,0.73,0.73}{##1}}}
\expandafter\def\csname PY@tok@kt\endcsname{\def\PY@tc##1{\textcolor[rgb]{0.69,0.00,0.25}{##1}}}
\expandafter\def\csname PY@tok@sc\endcsname{\def\PY@tc##1{\textcolor[rgb]{0.73,0.13,0.13}{##1}}}
\expandafter\def\csname PY@tok@sb\endcsname{\def\PY@tc##1{\textcolor[rgb]{0.73,0.13,0.13}{##1}}}
\expandafter\def\csname PY@tok@k\endcsname{\let\PY@bf=\textbf\def\PY@tc##1{\textcolor[rgb]{0.00,0.50,0.00}{##1}}}
\expandafter\def\csname PY@tok@se\endcsname{\let\PY@bf=\textbf\def\PY@tc##1{\textcolor[rgb]{0.73,0.40,0.13}{##1}}}
\expandafter\def\csname PY@tok@sd\endcsname{\let\PY@it=\textit\def\PY@tc##1{\textcolor[rgb]{0.73,0.13,0.13}{##1}}}

\def\PYZus{\char`\_}

\def\PYZgt{\char`\>}

\def\PYZhy{\char`\-}

\makeatother

\xdefinecolor{darkorange}{rgb}{0.8,0.5,0}
\definecolor{darkgreen}{rgb}{0,0.6,0}

\begin{document}
\title{Toward real-time data query systems in HEP}

\author{Jim Pivarski$^1$, David Lange$^1$, Thanat Jatuphattharachat$^2$}

\address{$^1$ Physics Department, Princeton University, Princeton, NJ 08544, USA}
\address{$^2$ Computer Engineering Chulalongkorn University, Krung Thep Maha Nakhon 10330, Thailand}

\ead{pivarski@princeton.edu}

\begin{abstract}
Exploratory data analysis tools must respond quickly to a user's questions, so that the answer to one question (e.g.\ a visualized histogram or fit) can influence the next. In some SQL-based query systems used in industry, even very large (petabyte) datasets can be summarized on a human timescale (seconds), employing techniques such as columnar data representation, caching, indexing, and code generation/JIT-compilation. This article describes progress toward realizing such a system for High Energy Physics (HEP), focusing on the intermediate problems of optimizing data access and calculations for ``query sized'' payloads, such as a single histogram or group of histograms, rather than large reconstruction or data-skimming jobs. These techniques include direct extraction of ROOT TBranches into Numpy arrays and compilation of Python analysis functions (rather than SQL) to be executed very quickly. We will also discuss the problem of caching and actively delivering jobs to worker nodes that have the necessary input data preloaded in cache. All of these pieces of the larger solution are available as standalone GitHub repositories, and could be used in current analyses.
\end{abstract}

\section{Benefits of a query service}

Despite everything that's changed in computing since the first AIHENP/ACAT workshop, one aspect of today's High Energy Physics (HEP) analyses that a physicist 27 years ago would recognize is the use of private skims. Rather than analyzing the reconstructed data directly, physicists skim it (dropping events) and slim it (dropping particle attributes), copy this summary closer to where they will be working, and then analyze the summary. The reason is performance: smaller data with fewer steps between the analyst and the data result in quicker exploratory plots, which allow the human analyst to be more engaged in the investigation. Long processing times are acceptable for established, push-button procedures, but a new analysis requires creative work and interactive discovery.

However, producing such a skim can take a long time--- weeks or months before seeing the first plot--- and is fraught with tradeoffs. Excluding important events or (more often) important particle attributes can be costly, and including unnecessary data only makes the process take longer and use more disk. Moreover, the copy uses extra resources (disk and processing), which will only get tighter as data volumes outpace computing budgets, and it can effectively ``price out'' small analysis groups. Since it is a distinct copy, it does not benefit from improvements or corrections to the source data until refreshed, and can often be a year or more out of date. Finally, it forces physicists to take time and attention away from physics and statistical issues to solve an IT problem. If the analysis can be performed without this skim, so much the better.

Real-time analysis on primary datasets is possible, and has been demonstrated in industry, usually in the form of large, distributed databases used internally by a company's data analytics team. These systems are built from Apache Drill, Impala, Kudu, Hawk, or Dremel and tuned to provide SQL access to petabytes of data in seconds. Unlike Hadoop and Spark, which are optimized for high throughput, these systems are optimized for low latency, and they use database-style indexes and query planning to avoid churning through all data with every request.

We can learn from these systems, adopting their techniques to provide similar access to HEP data, but there are some important differences between industry needs and HEP needs. A physicist's analysis is both simpler and more complex than a data scientist's analysis: it is simpler in that HEP does not require all-to-all processing, such as transposing a table of customer purchases to make product recommendations--- all of the information that is needed to process one physics event is contained in that event or small auxiliary tables, not all the other events. Once collected, a HEP dataset is immutable--- new events do not accumulate while the physicist is analyzing it. However, physics analyses are more complex in the depth of processing applied to that one event, examining many combinations of particles, searching for relationships. To help with the bookkeeping, physicists are inclined to think in terms of objects, iterating over collections of particle objects, rather than exploding and aggregating tables as in SQL.

Because of the differences, we do not expect an off-the-shelf database to solve our problem directly, but we can adapt established techniques to work for HEP.

\section{Optimizing for ``query sized'' payloads}

An analysis query is the smallest unit of exploratory analysis, in which the physicist asks one question, such as ``What is the distribution of muon $\eta$ for the second-highest $p_T$ muon in each event, subject to track-fit $\chi^2$ constraints?'' The conditions may be very complex, involving mass calculations, vertex fits, generated/reconstructed particle matching, computed with deeply nested loops and external function calls, but it is a request for one plot. The request and the plot are small enough to be transmitted through a slow network, and the data processing touches at most a dozen particle attributes out of thousands. Even a petabyte-sized dataset would only need to yield a terabyte of data, distributed among hundreds of workers, to service this request.

\begin{table}
\caption{\label{tab:10k} Rate of processing a ``query sized'' payload: filling one histogram of jet $p_T$ for all jets in a $t\bar{t}$ sample, illustrating the orders of magnitude lost to providing a full framework for heavy event processing (all single-threaded).}

\vspace{0.25 cm}
\begin{center}
\begin{tabular}{r l}
0.018~MHz & full framework (CMSSW, single-threaded C++) \\
0.029~MHz & load all 95 jet branches in ROOT \\
2.8~MHz & load jet $p_T$ branch (and no others) in ROOT \\
12~MHz & allocate C++ objects on heap, fill, delete \\
31~MHz & allocate C++ objects on stack, fill histogram \\
250~MHz & minimal ``for'' loop in memory (single-threaded C)
\end{tabular}
\end{center}
\vspace{-0.75 cm}
\end{table}

HEP frameworks are not designed for this style of data access. Table~\ref{tab:10k} shows how four orders of magnitude in processing rate are lost to provide all the services of a full event framework (CMSSW in this example). These include creating full C++ particle objects with all attributes loaded, allocated randomly on the heap for convenient memory management, compared to a sequential, contiguous, and possibly vectorized histogram fill over a simple array. These services are essential for ``heavy'' processing tasks like event reconstruction, which touch most attributes of each particle, but are inappropriate for ``query sized'' tasks like filling a histogram. Private skims strip most of this away for lightweight processing in a physicist's personal analysis environment; there is no reason a lightweight framework could not also be provided centrally.

Similar to low-latency databases used in industry, ROOT stores objects in an ``exploded'' form (called ``splitting'' in ROOT), with all values of each attribute in a separate array. This is even possible for hierarchically nested data, containing arbitrary length lists of objects, though multiple arrays with different lengths are required to encode it. This allows ROOT to only read attributes required for the analysis function--- such as a terabyte of a petabyte dataset.

However, databases like Apache Drill go one step further by leaving the data in their exploded form during processing, without materializing rows as runtime objects\cite{drill}. Selective reading accounts for the first two orders of magnitude in Table~\ref{tab:10k}, while eliminating object materialization accounts for most of the last two by skipping memory allocation, non-sequential memory scans, and permitting compiler optimizations like loop vectorization.

To permit this kind of access for ROOT data, Brian Bockelman and Zhe Zhang added a method to ROOT I/O that skips the {\tt GetEntry} (event materialization) process\cite{bulkio}. Query sized calculations on the resulting arrays (computing momentum magnitudes from components) run 5~times faster than a streamlined {\tt GetEntry} loop and 10~times faster than {\tt TTree::Draw} or {\tt TTreeReader} for uncompressed or LZ4-compressed input data.

\section{Code transformation and compilation}

For interactive analysis, it must be possible to define analysis functions at runtime and then compile them to run at full speed on arrays. Drill converts SQL to Java bytecode\cite{drill}, while Brian and Zhe's test cases use Numba to JIT-compile Python\cite{bulkio}. What these tools lack is a mechanism to convert a physicist's object oriented view of the data into array operations. Below, I will describe how we can do this and how the transformed code performs.

\begin{table}
\caption{\label{tab:exploding} Illustration of exploding nested, hierarchical objects. The data below are logically conceived as a list of lists of character/integer pairs, but stored in memory as four flat arrays describing the structure of the outer list, the inner list, the 1$^{\mbox{\scriptsize st}}$ and 2$^{\mbox{\scriptsize nd}}$ attributes.}

\begin{center}
\begin{tabular}{r l}
logical data & {\tt\small \textcolor{white}{[}\textcolor{blue}{[}\textcolor{violet}{[}(\textcolor{darkorange}{a},\textcolor{darkgreen}{1}), (\textcolor{darkorange}{b},\textcolor{darkgreen}{2}), (\textcolor{darkorange}{c},\textcolor{darkgreen}{3}), (\textcolor{darkorange}{d},\textcolor{darkgreen}{4})\textcolor{violet}{]}, \textcolor{violet}{[]}, \textcolor{violet}{[}(\textcolor{darkorange}{e},\textcolor{darkgreen}{5}), (\textcolor{darkorange}{f},\textcolor{darkgreen}{6})\textcolor{violet}{]}\textcolor{blue}{]}, \textcolor{blue}{[]}, \textcolor{blue}{[}\textcolor{violet}{[}(\textcolor{darkorange}{g},\textcolor{darkgreen}{7})\textcolor{violet}{]}\textcolor{blue}{]}\ \textcolor{white}{]}} \\\hline
outer offsets & {\tt\small \textcolor{blue}{[0,\ \ \ \ \ \ \ \ \ \ \ \ \ \ \ \ \ \ \ \ \ \ \ \ \ \ \ \ \ \ \ \ \ \ \ \ \ \ \ \ \ \ \ \ \ \ \ \ \ \ 3,\ \ 3,\ \ \ \ \ \ \ 4]}} \\
inner offsets & {\tt\small \textcolor{violet}{[\ 0,\ \ \ \ \ \ \ \ \ \ \ \ \ \ \ \ \ \ \ \ \ \ \ \ \ \ \ \ 4,\ \ 4,\ \ \ \ \ \ \ \ \ \ \ \ \ \ \ \ \ \ \ \ 6,\ \ \ \ \ \ 7]}} \\
1$^{\mbox{\scriptsize st}}$ attribute & {\tt\small \textcolor{darkorange}{[\ \ \ a,\ \ \ \ \ b,\ \ \ \ \ c,\ \ \ \ \ d,\ \ \ \ \ \ \ \ \ \ \ e,\ \ \ \ \ f,\ \ \ \ \ \ \ \ \ \ \ \ \ g\ \ \ \ \ \ ]}} \\
2$^{\mbox{\scriptsize nd}}$ attribute & {\tt\small \textcolor{darkgreen}{[\ \ \ \ \ 1,\ \ \ \ \ 2,\ \ \ \ \ 3,\ \ \ \ \ 4,\ \ \ \ \ \ \ \ \ \ \ 5,\ \ \ \ \ 6,\ \ \ \ \ \ \ \ \ \ \ \ \ 7\ \ \ \ ]}}
\end{tabular}
\end{center}
\vspace{-0.75 cm}
\end{table}

Given the exploded data shown in Table~\ref{tab:exploding} and user code such as
\begin{center}
\begin{minipage}{0.7\linewidth}
\begin{verbatim}
for outerlist in dataset:
   for innerlist in outerlist:
      for pair in innerlist:
         compute(pair.first, pair.second)
\end{verbatim}
\end{minipage}
\end{center}
we want to convert it to a form that doesn't reference ``{\tt dataset},'' ``{\tt outerlist},'' ``{\tt innerlist},'' or ``{\tt pair}'' as objects, instead deferring to ``{\tt outeroffsets},'' ``{\tt inneroffsets},'' ``{\tt first}'' and ``{\tt second},'' where the data are stored as arrays.

Such a transformation can be performed algorithmically on the user code's Abstract Syntax Tree (AST), converting the above into
\begin{center}
\begin{minipage}{0.7\linewidth}
\begin{verbatim}
for (i = 0;  i < 3;  i++)
   for (j = outeroffsets[i];  j < outeroffsets[i+1];  j++)
      for (k = inneroffsets[j];  k < inneroffsets[j+1];  k++)
         compute(first[k], second[k]);
\end{verbatim}
\end{minipage}
\end{center}
by replacing each ``{\tt outerlist}'' AST node with its corresponding ``{\tt outeroffsets[i]}'' and each ``{\tt pair.first}'' with its corresponding ``{\tt first[k]}.'' The first transformation rule eliminates references to list objects and the second eliminates references to record objects.

This transformation is like a type-inferring compilation pass, in which the types of dataset substructures must be propagated through the code, including assignments to new variables, and syntactic structures such as ``for'' loops have special transformation rules. Functions such as ``len'' (requesting the length of a list) must be overloaded as well (e.g.\ ``{\tt offsets[j+1] - offsets[j]}'').

In special cases such as this one, we can even convert the above into
\begin{center}
\begin{minipage}{0.7\linewidth}
\begin{verbatim}
for (k = 0;  k < inner[outer[3]];  k++)
   compute(first[k], second[k]);
\end{verbatim}
\end{minipage}
\end{center}
by recognizing that the outer and inner loops are total and sequential. Thus, when the transformed function is passed to a bytecode compiler such as Numba or Clang, the non-nested ``for'' loop may be more highly optimized, possibly vectorized (depending on ``{\tt compute}''). This code transformation technique is discussed in more detail in\cite{ieee}.

\begin{figure}
\includegraphics[width=0.47\linewidth]{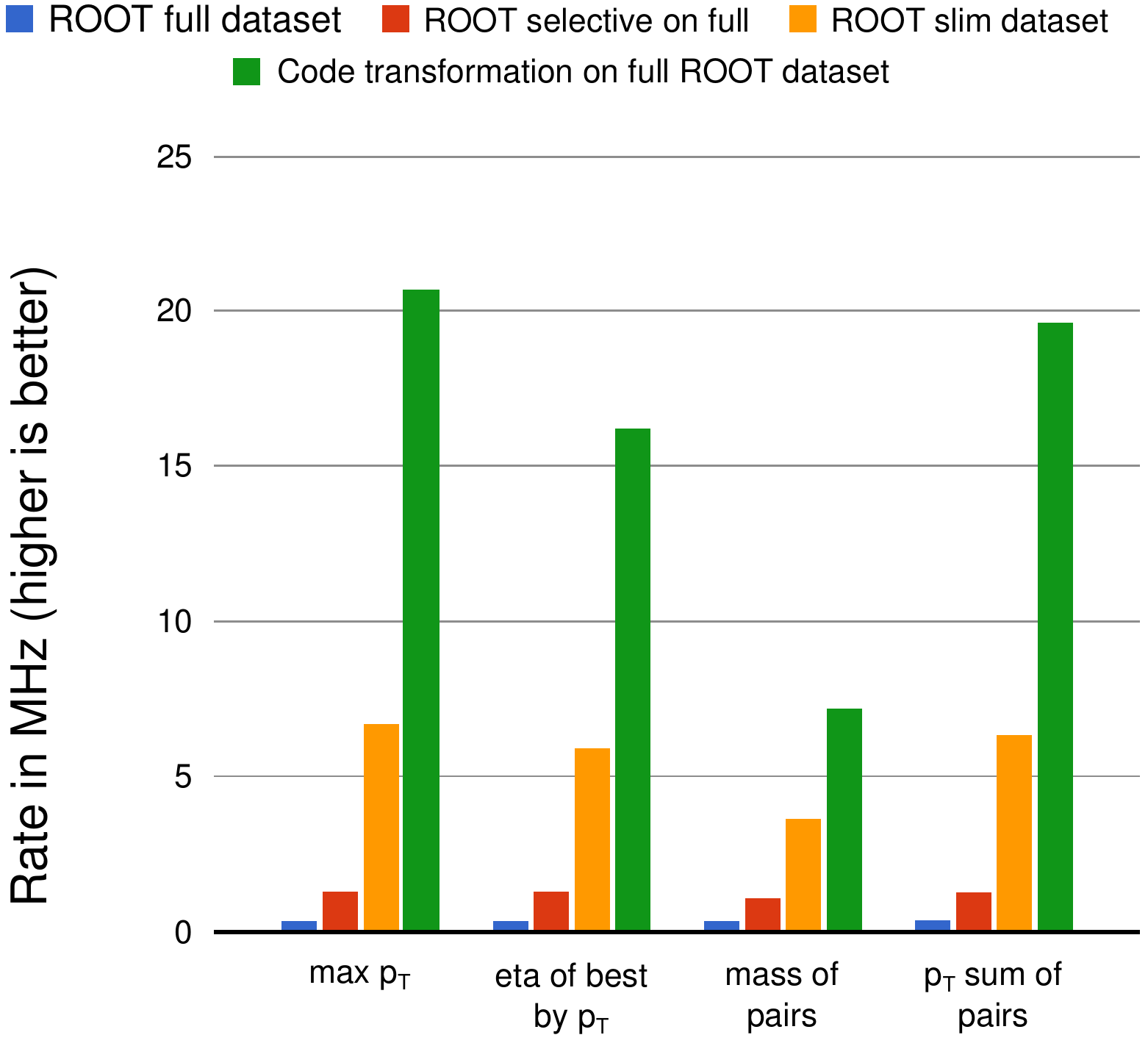} \hfill
\includegraphics[width=0.47\linewidth]{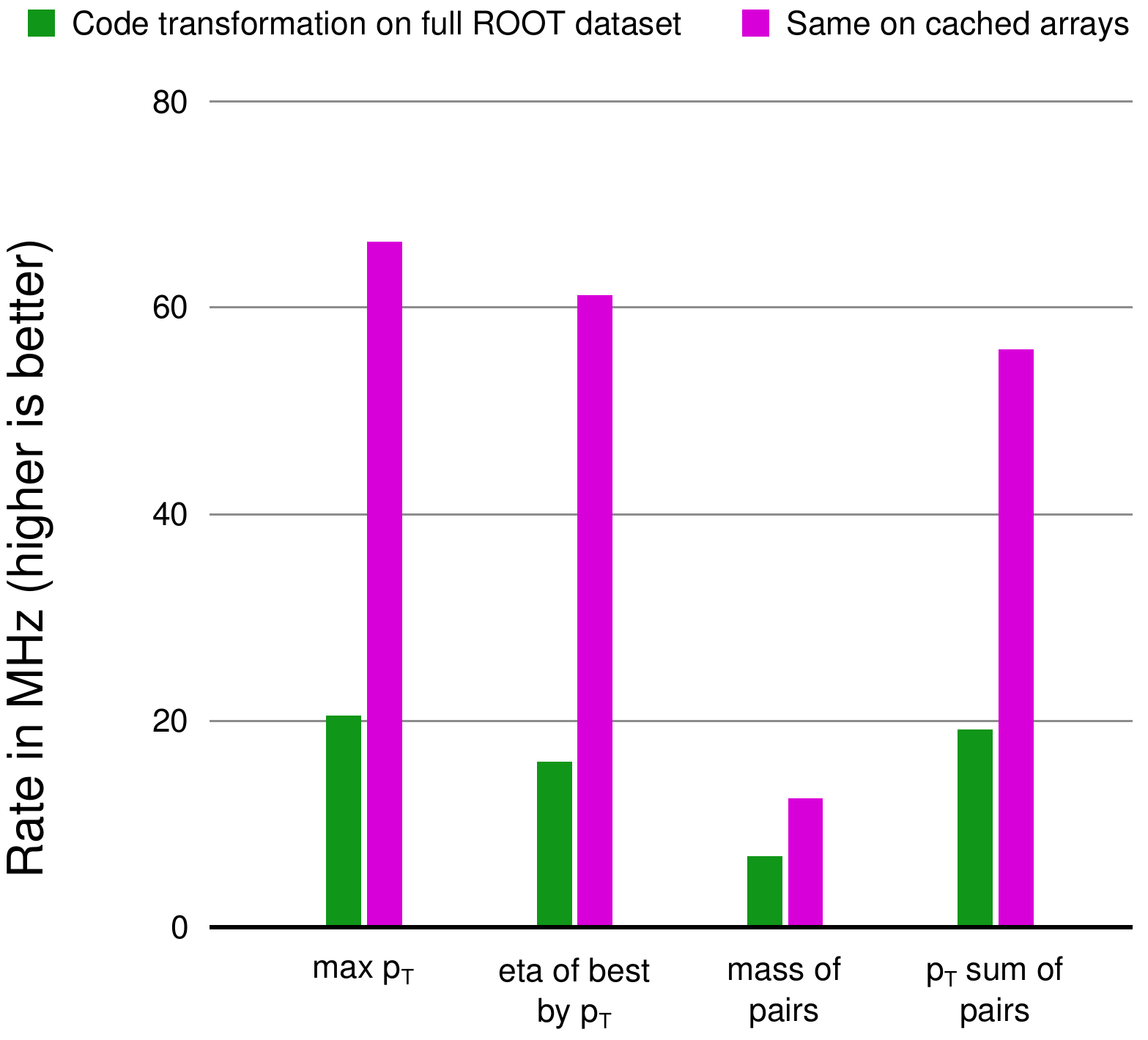}

\caption{\label{fig:codetrans} Processing rate of four analysis functions (Table~\ref{tab:code}) implemented in C++ with {\tt GetEntry} and our Python code transformation/compilation. ``ROOT full dataset'' loads all particle attributes, including unnecessary ones, ``selective on full'' uses selective reading ({\tt SetBranchStatus}), ``slim dataset'' uses a specially preselected input sample, ``Code transformation on full ROOT dataset'' uses Brian and Zhe's array extraction and our code transformation. Though uncompressed in warm cache, ROOT file reading is still the bottleneck; transformed code on raw arrays cached in memory is several times faster (all single-threaded).}
\end{figure}

The performance of the transformed code was studied with a simulated Drell-Yan dataset containing 5.4~million collisions in the CMS detector. Calculations were performed in single-threaded processes on a {\tt i2.xlarge} Amazon Web Services (AWS) instance, files were read from warmed cache, and the contents were uncompressed. Results are presented in Figure~\ref{fig:codetrans}. 

\begin{table}
\caption{\label{tab:code} Analysis functions tested in Figure~\ref{fig:codetrans}: ``max $p_T$'' is a per-event aggregation, ``eta of best'' adds the complication of maximizing one attribute while plotting another, ``mass of pairs'' iterates over distinct pairs of particles per event, computing an essential HEP function, and ``$p_T$ sum of pairs'' has the same loop structure without the expensive $\cosh$ and $\cos$ calls.}

\begin{center}
\begin{tabular}{p{0.21\linewidth} p{0.22\linewidth} p{0.24\linewidth} p{0.21\linewidth}}
\underline{{\small\bf max p$_{\mbox{\tiny T}}$}}
\begin{minipage}{\linewidth}\scriptsize
\begin{Verbatim}[commandchars=\\\{\}]
\PY{k}{for} \PY{n}{event} \PY{o+ow}{in} \PY{n}{dataset}{:}
  \PY{n}{maximum} \PY{o}{=} \PY{l+m+mf}{0.0}
  \PY{k}{for} \PY{n}{muon} \PY{o+ow}{in} \PY{n}{event}\PY{o}{.}\PY{n}{Muon}\PY{p}{:}
    \PY{k}{if} \PY{n}{muon}\PY{o}{.}\PY{n}{pt} \PY{o}{\PYZgt{}} \PY{n}{maximum}\PY{p}{:}
      \PY{n}{maximum} \PY{o}{=} \PY{n}{muon}\PY{o}{.}\PY{n}{pt}
  \PY{n}{fill\PYZus{}histogram}\PY{p}{(}\PY{n}{maximum}\PY{p}{)}
\end{Verbatim}

\end{minipage} &
\underline{{\small\bf eta of best by p$_{\mbox{\tiny T}}$}}
\begin{minipage}{\linewidth}\scriptsize
\begin{Verbatim}[commandchars=\\\{\}]
\PY{k}{for} \PY{n}{event} \PY{o+ow}{in} \PY{n}{dataset}{:}
  \PY{n}{maximum} \PY{o}{=} \PY{l+m+mf}{0.0}
  \PY{n}{best} \PY{o}{=} \PY{k}{None}
  \PY{k}{for} \PY{n}{muon} \PY{o+ow}{in} \PY{n}{event}\PY{o}{.}\PY{n}{muons}\PY{p}{:}
    \PY{k}{if} \PY{n}{muon}\PY{o}{.}\PY{n}{pt} \PY{o}{\PYZgt{}} \PY{n}{maximum}\PY{p}{:}
      \PY{n}{maximum} \PY{o}{=} \PY{n}{muon}\PY{o}{.}\PY{n}{pt}
      \PY{n}{best} \PY{o}{=} \PY{n}{muon}
  \PY{k}{if} \PY{n}{best} \PY{o+ow}{is not} \PY{k}{None}\PY{p}{:}
    \PY{n}{fill\PYZus{}histogram}\PY{p}{(}\PY{n}{best}\PY{o}{.}\PY{n}{eta}\PY{p}{)}
\end{Verbatim}

\end{minipage} &
\underline{{\small\bf mass of pairs}}
\begin{minipage}{\linewidth}\scriptsize
\begin{Verbatim}[commandchars=\\\{\}]
\PY{k}{for} \PY{n}{event} \PY{o+ow}{in} \PY{n}{dataset}{:}
  \PY{n}{n} \PY{o}{=} \PY{n+nb}{len}\PY{p}{(}\PY{n}{event}\PY{o}{.}\PY{n}{muons}\PY{p}{)}
  \PY{k}{for} \PY{n}{i} \PY{o+ow}{in} \PY{n+nb}{range}\PY{p}{(}\PY{n}{n}\PY{p}{)}\PY{p}{:}
    \PY{k}{for} \PY{n}{j} \PY{o+ow}{in} \PY{n+nb}{range}\PY{p}{(}\PY{n}{i}\PY{o}{+}\PY{l+m+mi}{1}\PY{p}{,} \PY{n}{n}\PY{p}{)}\PY{p}{:}
      \PY{n}{m1} \PY{o}{=} \PY{n}{event}\PY{o}{.}\PY{n}{muons}\PY{p}{[}\PY{n}{i}\PY{p}{]}
      \PY{n}{m2} \PY{o}{=} \PY{n}{event}\PY{o}{.}\PY{n}{muons}\PY{p}{[}\PY{n}{j}\PY{p}{]}
      \PY{n}{mass} \PY{o}{=} \PY{n}{sqrt}\PY{p}{(}
        \PY{l+m+mi}{2}\PY{o}{*}\PY{n}{m1}\PY{o}{.}\PY{n}{pt}\PY{o}{*}\PY{n}{m2}\PY{o}{.}\PY{n}{pt}\PY{o}{*}\PY{p}{(}
        \PY{n}{cosh}\PY{p}{(}\PY{n}{m1}\PY{o}{.}\PY{n}{eta} \PY{o}{\PYZhy{}} \PY{n}{m2}\PY{o}{.}\PY{n}{eta}\PY{p}{)} \PY{o}{\PYZhy{}}
        \PY{n}{cos}\PY{p}{(}\PY{n}{m1}\PY{o}{.}\PY{n}{phi} \PY{o}{\PYZhy{}} \PY{n}{m2}\PY{o}{.}\PY{n}{phi}\PY{p}{)}\PY{p}{)}\PY{p}{)}
      \PY{n}{fill\PYZus{}histogram}\PY{p}{(}\PY{n}{mass}\PY{p}{)}
\end{Verbatim}

\end{minipage} &
\underline{{\small\bf p$_{\mbox{\tiny T}}$ sum of pairs}}
\begin{minipage}{\linewidth}\scriptsize
\begin{Verbatim}[commandchars=\\\{\}]
\PY{k}{for} \PY{n}{event} \PY{o+ow}{in} \PY{n}{dataset}{:}
  \PY{n}{n} \PY{o}{=} \PY{n+nb}{len}\PY{p}{(}\PY{n}{event}\PY{o}{.}\PY{n}{muons}\PY{p}{)}
  \PY{k}{for} \PY{n}{i} \PY{o+ow}{in} \PY{n+nb}{range}\PY{p}{(}\PY{n}{n}\PY{p}{)}\PY{p}{:}
    \PY{k}{for} \PY{n}{j} \PY{o+ow}{in} \PY{n+nb}{range}\PY{p}{(}\PY{n}{i}\PY{o}{+}\PY{l+m+mi}{1}\PY{p}{,} \PY{n}{n}\PY{p}{)}\PY{p}{:}
      \PY{n}{m1} \PY{o}{=} \PY{n}{event}\PY{o}{.}\PY{n}{muons}\PY{p}{[}\PY{n}{i}\PY{p}{]}
      \PY{n}{m2} \PY{o}{=} \PY{n}{event}\PY{o}{.}\PY{n}{muons}\PY{p}{[}\PY{n}{j}\PY{p}{]}
      \PY{n}{s} \PY{o}{=} \PY{n}{m1}\PY{o}{.}\PY{n}{pt} \PY{o}{+} \PY{n}{m2}\PY{o}{.}\PY{n}{pt}
      \PY{n}{fill\PYZus{}histogram}\PY{p}{(}\PY{n}{s}\PY{p}{)}
\end{Verbatim}

\end{minipage}
\end{tabular}
\end{center}
\vspace{-0.75 cm}
\end{table}

\section{Distributed processing with cache}

The optimizations discussed above focus on increasing single-threaded throughput, but of course the main benefit of a centralized query service is parallelization. If we attain our latency goal of no more than a second per plot and a hundred physicists are online, submitting a query every ten seconds, then each physicist would get a tenth of the whole cluster at a time.

HEP analysis functions are embarrasingly parallel, which would scale linearly if data can be delivered to them efficiently. However, this depends critically on caching, since data processing is considerably faster than disk access, and caching is local. An input dataset in memory on one machine is only useful if subsequent jobs requiring that input are sent to the same machine.

\begin{figure}[b]
\begin{center}
\includegraphics[width=0.5\linewidth]{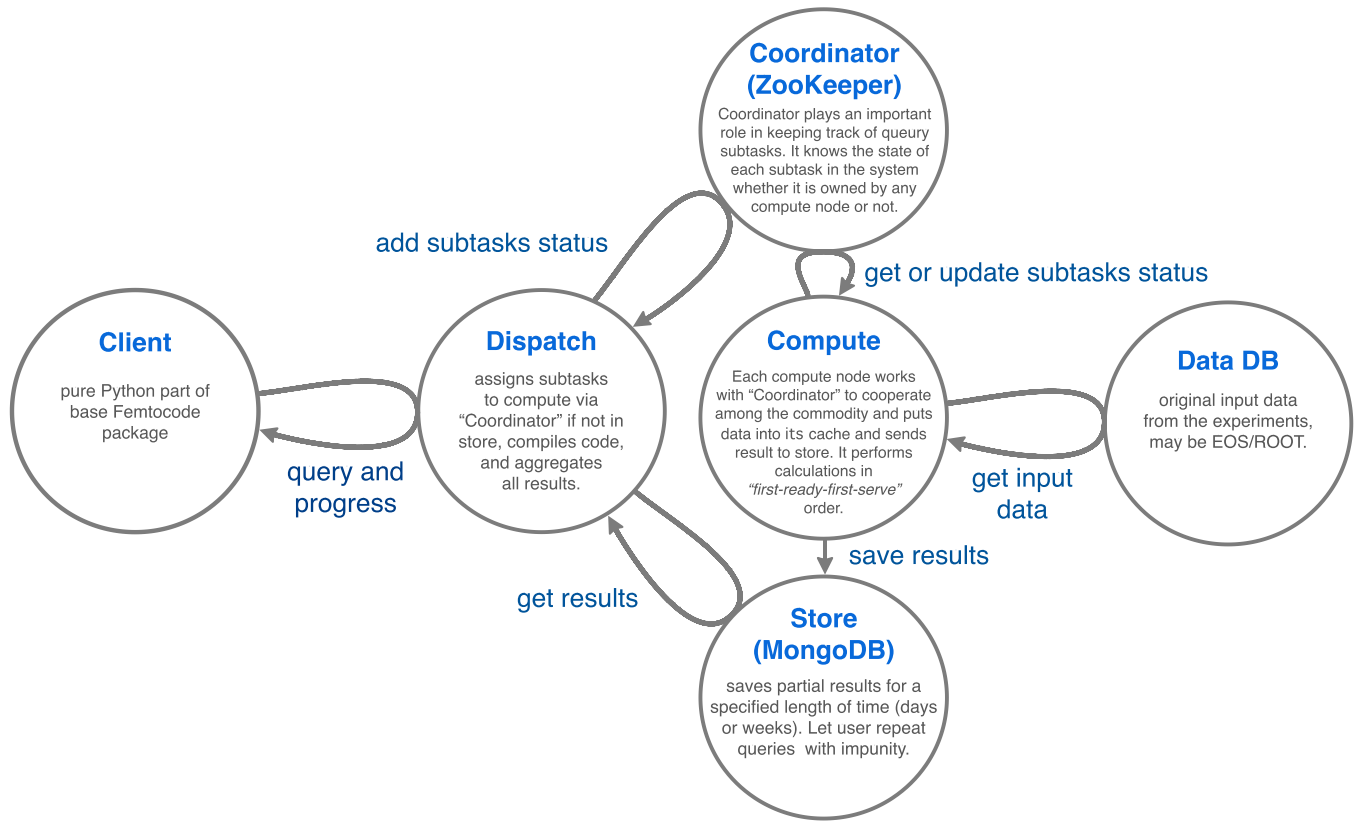}
\end{center}

\caption{\label{fig:distributed} Schematic for distributed query processing to minimize cache misses (see text).}
\end{figure}

Rather than dispatch subtasks round-robin or to the least busy compute node, we want compute nodes to pull subtasks with a preference for input data they already have in cache. Figure~\ref{fig:distributed} shows a scheme for accomplishing this, using Apache Zookeeper to advertise new subtasks and globally mark them as in progress and delete them when done. Compute nodes pull subtasks rather than having work pushed to them because only the compute nodes know which input datasets are locally in cache. Pulling work also reduces the effect of stragglers for the same reason as work-stealing. Additionally, the work-pulling algorithm should not strictly require input data to be in cache, but should check for work in two rounds: the first takes only cache-local work, but if there is no cache-local work to do, compute nodes will take any work after a sub-second delay. This allows the compute nodes that can best service the request to have first dibs, but if there's too much for them to do, new compute nodes will download the input data so that the handling of a popular dataset elastically scales with demand.

For the most common type of query, histogram aggregation, subtasks represent partial histograms that must be aggregated centrally. For responsiveness, we imagine storing partial histograms in a document database like MongoDB and aggregating whatever is available at regular intervals. That way, the user would see results accumulate interactively and can cancel malformed queries. The range of supported tasks can be extended by adopting generalized aggregation with Histogrammar\cite{histogrammar}.

This system was developed in draft form by one of us (Thanat)\cite{thanat}, to serve as a model for development. This draft system lacks only realistic work payloads, substituting them with random delays.

\section{Standalone pieces and the big picture}

A centralized query service that is as fast and convenient as private skims would have a profound impact on the working habits of thousands of physicists. It doesn't represent an incremental change of existing infrastructures, and as such, needs quite a few new components to make it work.

Rather than trying to introduce this system in a ``big bang,'' we are rolling out individual components, even if they tend to increase reliance on private skims in the short term. For instance, array access to ROOT data and Python code transformation/compilation can accelerate and simplify analysis using skims. We are considering ways of implementing a distributed system as a virtual ROOT file, to provide physicists with a familiar interface, even if it isn't used with the accelerated query engine. This way, we can get feedback about parts of the system before the whole is in place. This method also encourages spin-off technologies.

The following software is either available now or soon.
\begin{itemize}
\item Array access to ROOT data, via Numpy and {\tt TTreeReaderFast}, is scheduled for {\bf ROOT 6.14}\cite{root}. Users willing to compile ROOT may try the feature early using a custom fork\cite{root-brian}.
\item The same interface is available now in a Python package called {\bf uproot}\cite{uproot}.
\item The code transformation techniques described here are implemented in {\bf OAMap}\cite{oamap}, which accelerates functions on data in Apache Arrow\cite{arrow} format, with on-the-fly conversion from ROOT to Arrow. The purpose of targeting Arrow rather than ROOT directly is to attract contributions from Big Data projects that rely on Arrow.
\end{itemize}

Readers are encouraged to try out these packages and report any usability issues, as this will guide the process of developing the query system.

\section*{Acknowledgements}

This work was supported by the National Science Foundation under grants ACI-1450377 and PHY-1624356.

\section*{References}

\bibliographystyle{IEEEtran}
\bibliography{mybibliography}

\begin{thebibliography}{10}
\providecommand{\url}[1]{#1}
\csname url@samestyle\endcsname
\providecommand{\newblock}{\relax}
\providecommand{\bibinfo}[2]{#2}
\providecommand{\BIBentrySTDinterwordspacing}{\spaceskip=0pt\relax}
\providecommand{\BIBentryALTinterwordstretchfactor}{4}
\providecommand{\BIBentryALTinterwordspacing}{\spaceskip=\fontdimen2\font plus
\BIBentryALTinterwordstretchfactor\fontdimen3\font minus
  \fontdimen4\font\relax}
\providecommand{\BIBforeignlanguage}[2]{{%
\expandafter\ifx\csname l@#1\endcsname\relax
\typeout{** WARNING: IEEEtran.bst: No hyphenation pattern has been}%
\typeout{** loaded for the language `#1'. Using the pattern for}%
\typeout{** the default language instead.}%
\else
\language=\csname l@#1\endcsname
\fi
#2}}
\providecommand{\BIBdecl}{\relax}
\BIBdecl

\bibitem{drill}
\BIBentryALTinterwordspacing
M.~Hausenblas and J.~Nadeau, ``Apache drill: Interactive ad-hoc analysis at
  scale,'' \emph{Big Data}, vol. 1(2), pp. 100--104, 2013. [Online]. Available:
  \url{https://doi.org/10.1089/big.2013.0011}
\BIBentrySTDinterwordspacing

\bibitem{bulkio}
\BIBentryALTinterwordspacing
{Brian Bockelman, Zhe Zhang, and Jim Pivarski}, ``{ROOT-BulkIO and Numpy
  interface},'' \emph{Journal of Physics: Conference Series}, 2017. [Online].
  Available: \url{https://indico.cern.ch/event/567550/contributions/2627167/}
\BIBentrySTDinterwordspacing

\bibitem{ieee}
J.~{Pivarski}, P.~{Elmer}, B.~{Bockelman}, and Z.~{Zhang}, ``{Fast Access to
  Columnar, Hierarchical Data via Code Transformation},'' \emph{ArXiv
  e-prints}, Aug. 2017.

\bibitem{histogrammar}
\BIBentryALTinterwordspacing
J.~Pivarski, A.~Svyatkovskiy, F.~Schenck, and B.~Engels, ``histogrammar-python:
  1.0.0,'' Sep. 2016. [Online]. Available:
  \url{https://doi.org/10.5281/zenodo.61418}
\BIBentrySTDinterwordspacing

\bibitem{thanat}
\BIBentryALTinterwordspacing
{Thanat Jatuphattharachat}. (2017) {femto-mesos}. [Online]. Available:
  \url{https://github.com/JThanat/femto-mesos}
\BIBentrySTDinterwordspacing

\bibitem{root}
R.~Brun and F.~Rademakers, ``{ROOT: An object oriented data analysis
  framework},'' \emph{Nucl. Instrum. Meth.}, vol. A389, pp. 81--86, 1997.

\bibitem{root-brian}
\BIBentryALTinterwordspacing
{Brian Bockelman}. (2017) {root-bulkapi-fastread-v2}. [Online]. Available:
  \url{http://github.com/bbockelm/root/tree/root-bulkapi-fastread-v2}
\BIBentrySTDinterwordspacing

\bibitem{uproot}
\BIBentryALTinterwordspacing
{Jim Pivarski}. (2017) {``Uproot: Minimalist ROOT I/O in pure Python and
  Numpy.''}. [Online]. Available: \url{https://github.com/scikit-hep/uproot}
\BIBentrySTDinterwordspacing

\bibitem{oamap}
\BIBentryALTinterwordspacing
------. (2017) {``OAMap: Toolset for Computing Directly on Hierarchically
  Nested, Columnar Data''}. [Online]. Available:
  \url{https://github.com/diana-hep/oamap}
\BIBentrySTDinterwordspacing

\bibitem{arrow}
\BIBentryALTinterwordspacing
{The Apache Arrow team}. (2016) {``Apache Arrow: Powering Columnar In-Memory
  Analytics''}. [Online]. Available: \url{https://arrow.apache.org/}
\BIBentrySTDinterwordspacing

\end{thebibliography}

\end{document}